\documentclass[12pt,draftcls,onecolumn]{IEEEtran}
\usepackage{graphicx}
\usepackage{subfigure}

\usepackage{cite}
\usepackage{url}

\usepackage{algorithm}
\usepackage{algorithmic}
\usepackage{amsmath, amssymb,amsthm}
\usepackage{balance}
\usepackage{capt-of}

\begin{document}

%don't want date printed
%\date{}

\title{Heterogeneous Coexistence of Cognitive Radio Networks in TV White Space\thanks{A preliminary version of portions of this material has appeared in~\cite{globecom_13}.}}

\author{
\small
Kaigui Bian$^\star$, Lin Chen$^{\star,\sharp}$, Yuanxing Zhang$^\star$, Jung-Min ``Jerry" Park$^\dagger$, Xiaojiang Du$^\ddagger$, and Xiaoming Li$^\star$\\
$^\star$Institute of Network Computing and Information System, School of EECS, Peking University, Beijing, China\\
$^\sharp$Department of Electrical Engineering, Yale University, New Haven, CT, USA\\
$^\dagger$Bradley Department of Electrical and Computer Engineering, Virginia Tech, Blacksburg, VA, USA\\
$^\ddagger$Department of Computer and Information Sciences, Temple University, Philadelphia, PA, USA\\
\{bkg,abratchen,longo,lxm\}@pku.edu.cn$^\star$, lin.chen@yale.edu$^\sharp$, jungmin@vt.edu$^\dagger$, dux@temple.edu$^\ddagger$\\
}

\maketitle

% Use the following at camera-ready time to suppress page numbers.
% Comment it out when you first submit the paper for review.
\thispagestyle{plain}

%%%%%%%%%%%%%%%%%%%%%%%%%%%%%%%%%%%%%%%%%%%%%%%%%
%%%%%%%%%%% Response to comments
%%%%%%%%%%%%%%%%%%%%%%%%%%%%%%%%%%%%%%%%%%%%%%%%%
% 1. comparison to related work: no related work on heter coexistence scheme. FSSE or our defined index? The later is better.
% 2. applicable to both CSMA and TDMA. LBT = CSMA, and Random channel selection are not suitable words in texts.
% 3. our focus: channel allocation, not channel sharing (only direct coordination)
% 4. no vertical coexistence, but only horizontal coexistence, like self-coexistence in Bo's paper.
% 4. Define complex variables.

\begin{abstract}
% new wireless technologies over TV white space
Wireless standards (e.g., IEEE 802.11af and 802.22) have been developed for enabling opportunistic access in TV white space (TVWS) using cognitive radio (CR) technology.
% heterogeneous coexistence problem is important
When heterogeneous CR networks that are based on different wireless standards operate in the same TVWS, coexistence issues can potentially cause major problems.
% challenges
Enabling collaborative coexistence via \emph{direct} coordination between heterogeneous CR networks is very challenging, due to incompatible MAC/PHY designs of coexisting networks, requirement of an over-the-air common control channel for inter-network communications, and time synchronization across devices from different networks. Moreover, such a coexistence scheme would require competing networks or service providers to exchange sensitive control information that may raise conflict of interest issues and customer privacy concerns.
% Contributions: ecology-inspired coexistence via indirect coordination
In this paper, we present an architecture for enabling collaborative coexistence of heterogeneous CR networks over TVWS, called \emph{Symbiotic Heterogeneous coexistence ARchitecturE (SHARE)}.
% Define "indirect coordination" first before using it. Because coexistence cannot avoid coordination
By mimicking the symbiotic relationships between heterogeneous organisms in a stable ecosystem, SHARE establishes an \emph{indirect} coordination mechanism between heterogeneous CR networks via a \emph{mediator} system, which avoids the drawbacks of direct coordination.
% Main theories
SHARE includes two spectrum sharing algorithms whose designs were inspired by well-known models and theories from theoretical ecology, viz, the interspecific competition model and the ideal free distribution model.
\end{abstract}

\section{Introduction}\label{sec:intro}
% benefits of TV white space
The TV ``white space'' (TVWS) has the potential of providing significant bandwidth in frequencies that have very favorable propagation characteristics (i.e., long transmission ranges and superior capability of penetrating objects)~\cite{whitefi_sigcomm09}.
In the U.S., United Kingdom, and other countries, changes in the regulatory rules have been made or are being amended to open up the TVWS for opportunistic operations of unlicensed (or secondary) users on a non-interference basis to licensed users (a.k.a. incumbent or primary users)~\cite{FCC_openTV}. Industry and research stakeholders have initialized standardization efforts to enable the utilization of TVWS by leveraging \emph{cognitive radio} (CR) technology. These efforts include IEEE 802.22 Wireless Regional Area Networks (WRAN)~\cite{80222}, IEEE 802.11af (WiFi over TVWS)~\cite{80211af}, ECMA 392 (WPAN over TVWS)~\cite{Ecma392}, etc. All of these standards rely on CR technology to overcome the challenging interference issues between incumbent and secondary networks as well as between secondary networks. In this paper, we simply use the term ``CR network'' to denote a CR-enabled wireless network operating over TVWS.

% Taxonomy of coexistence: we focus on heterogeneous coexistence
The coexistence of secondary wireless networks in TVWS can be broadly classified into two categories \cite{tccn2016}: \emph{heterogeneous coexistence} and \emph{homogeneous coexistence} (a.k.a. self coexistence). The former refers to the coexistence of networks that employ different wireless technologies (e.g., WiFi and Bluetooth) and the latter refers to the coexistence of networks that employ the same wireless technology (e.g., neighboring 802.22 networks). There are two types of coexistence schemes: \emph{non-collaborative} and \emph{collaborative} coexistence schemes.
\begin{itemize}
  \item A non-collaborative coexistence scheme is the only feasible approach when there are no means of coordination between the coexisting networks. In the existing literature, such an approach has been used to address the heterogeneous coexistence of WiFi and ZigBee networks \cite{wifi_zigbee,zifi} as well as the homogeneous coexistence of uncoordinated WiFi deployments \cite{CH_Mobicom06} and femto cell deployments \cite{femto_MobiHoc09,femto_Mobicom10}.
  \item A collaborative coexistence scheme can be employed when coexisting networks can directly coordinate their operations, and examples of such an approach include coexistence schemes for cellular networks \cite{cellular_ICNP10,cellular_Mobicom06} and 802.22 networks \cite{8022_TVT_game,80222_globe_game,80222_globe08,80222_odsc}. The recently formed IEEE 802.19.1 task group (TG) was chartered with the task of developing standardized methods, which are radio access technology-independent, for enabling coexistence among dissimilar or independently operated wireless networks~\cite{802191}. This standard is currently being developed, and it has yet to prescribe solid solutions.
\end{itemize}
As described below, existing coexistence schemes---both non-collaborative and collaborative---\emph{cannot} adequately address the problems posed by the coexistence of heterogeneous CR networks \cite{ukc2006}.

Non-collaborative coexistence schemes are simpler and cheaper to deploy, but not as effective as collaborative schemes. Moreover, non-collaborative schemes cannot facilitate the coexistence among networks with incompatible MAC strategies (e.g., coexistence between contention-based and reservation-based MAC protocols) and cannot adequately address the hidden node problem.

Collaborative strategies are stricken with a number of very difficult technical and policy problems. First, coexisting networks would need to exchange spectrum sharing control information over a \emph{common control channel}, and the realization of such a channel may require a broad standardization effort across secondary systems that would be costly. Second, even if an effective means of inter-network communications exists, implementation of collaborative strategies would rely on \emph{time synchronization} across devices from different networks. Achieving synchronization over a potentially large number of coexisting TVWS networks may not be feasible. Third, collaborative approaches would require the coexisting networks to exchange potentially sensitive information---such as traffic load, bandwidth requirements, and network characteristics---to negotiate partitioning of the spectrum. Exchanging such information between competing wireless networks or service providers could potentially raise conflict-of-interest issues and customer privacy concerns. Hence, it is difficult to find a global or centralized decision maker to allocate spectrum for all competing networks.

% design requirements of heter coexistence
% 1. no centralized decision maker
% 2. no clock synchronization
% 3. no direct coordination---no pairwise sharing of control information

In this paper, we propose a coexistence framework, called the \emph{Symbiotic Heterogeneous coexistence ARchitectuRE} (\emph{SHARE}), for enabling collaborative coexistence among heterogeneous CR networks. As its name implies, the proposed framework was inspired by the inter-species relations that exist in biological ecosystems. A \emph{symbiotic} relation is a term used in biology to describe the coexistence of different species that form relations via indirect coordination. SHARE exploits a \emph{mediator} system (e.g., the 802.19.1 system) to establish the \emph{indirect} coordination mechanism between coexisting networks.

In SHARE, the heterogeneous coexistence problem is addressed in two ways. First, we propose an \emph{ecology-inspired spectrum allocation} algorithm inspired by an interspecific resource competition model. This algorithm enables a CR network to calculate the amount of spectrum that it should appropriate without direct negotiation with competing networks. Second, we propose a \emph{foraging-based channel selection} algorithm, inspired by the ideal free distribution model in the optimal foraging theory, that enables each CR network to select the most appropriate TVWS channels. Note that these algorithms do not require coexisting networks to engage in direct negotiation. Our analytical and simulation results show that SHARE guarantees weighted-fairness in partitioning spectrum and improves spectrum utilization.

The rest of this paper is organized as follows. We provide background knowledge of the mediator system, and theoretical ecology in Section~\ref{sec:bkg}. In Section~\ref{sec:overview}, we give an overview of SHARE. We present the two SHARE algorithms and provide analytical results in Sections~\ref{sec:LV} and \ref{sec:foraging}, respectively. In Section~\ref{sec:sim}, we evaluate the performance of SHARE using the simulation. We conclude the paper in Section~\ref{sec:conclude}.

\section{Technical Background}\label{sec:bkg}
As stated previously, SHARE employs a mediator system to establish an indirect coordination mechanism between CR networks. Due to the conflict-of-interest issues and customer privacy concerns between competing networks, the mediator is not the global decision maker, and it only forwards sanitized data to the coexisting networks. Using the forwarded information, each CR network makes coexistence decisions autonomously using the two algorithms proposed in this paper.

\subsection{The Mediator System}
\begin{figure}[t]
\centering
\includegraphics[width=3in]{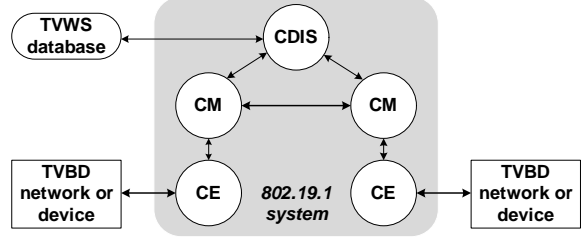}
%\vspace{-0.1in}
\caption {IEEE 802.19.1 system architecture.} \label{fig_80219}
\end{figure}

The IEEE 802.19.1 system is a good candidate to serve as the mediator. The IEEE 802.19.1 system \cite{802191} defines a set of logical entities and a set of standardized interfaces for enabling coordination between heterogeneous CR networks. In Figure~\ref{fig_80219}, we show the architecture of an 802.19.1 system which includes three entities in the grey box: (1) the coexistence manager (CM) acts as the local decision maker of the coexistence process; (2) the coexistence database and information server (CDIS) provides coexistence-related control information to the CMs, and (3) the coexistence enabler (CE) enables communications between the 802.19.1 system and the TV band device (TVBD) network. The TVWS database indicates the list of channels used by incumbent users and their locations, and it is connected to the 802.19.1 system via backhaul connections.

\subsection{Coexistence-related Constructs in Ecology}
In this subsection, we review the models and constructs in theoretical ecology that inspired the design of SHARE.

\subsubsection{Interspecific competition model}
In ecology, interspecific competition is a distributed form of competition in which individuals of different species compete for the same resource in an ecosystem without direct interactions between them~\cite{book_coexistence}. The impact of interspecific competition on populations have been formalized in a mathematical model called the Lotka-Volterra (L-V) competition model~\cite{LV_model1,LV_model2}. In this model, the impact on population dynamics of species $i$ can be calculated separately by a differential equation given below:
\begin{equation}\label{eqt_LV_Ni}
\frac{d N_i}{d t} = r_i N_i \left ( 1 - \frac{ N_i + \sum_{j\neq i} \alpha_{ij} N_j  } {K_i} \right).
\end{equation}
In this equation, $N_i$ is the population size of species $i$, $K_i$ is the carrying capacity (which is the maximum population of species $i$ if it is the only species present in the environment), $r_i$ is the intrinsic rate of increase, and $\alpha_{ij}$ is the competition coefficient which represents the impact of species $j$'s population growth on the population dynamics of species $i$.

\subsubsection{Ideal free distribution (IFD)}
In the optimal foraging theory (OFT), animals forage in such a way as to maximize their net energy intake per unit time, and the ideal free distribution (IFD) was introduced in~\cite{IFD} as an OFT model. IFD has been used to analyze how animals distribute themselves across different patches of resources. Suppose there are a number of disjoint patches of resource (e.g., food) to be allocated to animals in a given environment. These patches are indexed by $i = 0, . . . , p-1$. Let $x_i$ denote the amount of animals in the $i$-th patch. The total population of animals in the environment is $\rho = \sum_{i\in [0, p-1]} x_i$.

% suitability
Let $u_i$ be the \emph{suitability} of the $i$-th patch, which quantifies the patch's attractiveness to the animals.
\begin{equation}\label{eqt_suitability}
u_i = \frac{a_i}{x_i},
\end{equation}
where $a_i$ represents the nutrients per second in patch~$i$.

The \emph{fitness} of an animal in a patch is typically assumed to be equal to the suitability of the patch. Let $\phi_i$ be the fitness of an animal in the $i$-th patch, and thus $\phi_i = u_i$.

% How IFD is achieved
The IFD is a sequential allocation process where more (less) animals are distributed in patches with higher (lower) suitability. The IFD's \emph{equilibrium} point is achieved when each animal simultaneously maximizes its own fitness by moving into the patch with the highest suitability. At the equilibrium point, the suitability of all patches and the fitness of all animals equalize.

% matching rule
The ``input matching rule'' is used to characterize the equilibrium point of an IFD process~\cite{matching_rule}, and is prescribed as follows: animals are distributed such that for all $i \in [0, p-1]$,
\begin{equation}\label{eqt_matching_rule}
\frac{x_i}{\sum_{j=0}^{p-1} x_j} = \frac{a_i}{\sum_{j=0}^{p-1} a_j}.
\end{equation}\label{eqt_matching_rule}

\section{Overview of SHARE}\label{sec:overview}
In this section, we present the system model, underlying assumptions and the architecture of SHARE.

\subsection{System Model}
We assume $n$ heterogeneous CR networks are co-located, and they coexist in the same TVWS that includes $N$ TVWS channels with identical bandwidth. Let $\mathcal{K}$ denote this set of CR networks, and all of these networks in $\mathcal{K}$ are registered with the mediator system.
%We label each of these networks as network~$i$, where $i\in[1,n]$.
Every CR network is composed of multiple TVBDs and a CR-enabled base station (BS) (or access point). The TVWS channels are labeled with indices $0, 1, ..., N-1$.

\textbf{Number of TVWS channels.} In this paper, we focus on the case when the number of channels $N$ is no less than the number of co-located heterogeneous networks, $n$. That is, every network is allowed to exclusively occupy at least one channel. Here, we assume the 6 MHz TVWS channel.
% When $N < n$, some co-located networks have to operate on the same channel, and channel sharing techniques are needed to avoid the inter-network interference. However, channel sharing requires tight synchronization and exchange of sensitive control information (e.g., individual networks' traffic schedules) between any two co-located networks that may have the conflict of interest issues and customer privacy concerns. Hence, these assumptions are difficult to assume in heterogeneous coexistence scenarios. << Remove >>

\textbf{The bandwidth requirement.} We define the \emph{bandwidth requirement} of a CR network as the number of TVWS channels that it needs to satisfy the QoS requirements of its traffic load. Let $R_i$ denote the bandwidth requirement of network~$i$.
% << Here, are you assuming 6 MHz-wide channels? Specify this. >>

% define indirect coordination mechanism to finish the tasks
\textbf{The mediator-based indirect coordination.} SHARE establishes a mediator-based \emph{indirect} coordination mechanism between coexisting CR networks. There is no direct coordination between the coexisting networks, and they have to interact with each other by exchanging control information through a third-party mediator. Specifically, SHARE utilizes a CDIS (which is one of the components of an 802.19.1 system) as a mediator. Note that CDIS is not a global or centralized decision maker, but rather it is an information directory server with simple data processing capabilities.

\textbf{Exchange of sanitized information.} The mediator helps address conflict-of-interest issues and customer privacy concerns, which may arise when coexisting networks operated by competing service providers are required to exchange sensitive traffic information in order to carry out coexistence mechanisms. If needed, the mediator \emph{sanitizes} the sensitive information received from the coexisting networks and then returns the sanitized information back to them. The coexisting networks execute their coordinated coexistence mechanisms using the sanitized data.

\begin{table}
% \hspace{0.033\textwidth}
  \centering
   \caption{\textnormal{A mapping between biological and CR network ecosystems.}}
  \label{tab_mapping}
%\small
\begin{tabular}{|c||c|}
\hline
\bfseries      \textbf{Biological ecosystem}   &  \textbf{CR network system}   \\
\hline\hline
A species                               & A CR network          \\
\hline
Population                              &   Spectrum share           \\
  of a species                          &    of a CR network             \\
\hline
Population dynamics                       &      Dynamics of         \\
(growth or decline)                       &   spectrum share           \\
\hline
\end{tabular}
%\normalsize
\end{table}

% for notation consistency: $X$ = a value; $\mathcal{X}$ = a set; $\mathbf{X}$ = a vector
\subsection{The Two Tasks of Spectrum Sharing}
In a spectrum sharing process, a CR network has to perform two tasks: (1) figure out how much spectrum it can appropriate given its bandwidth requirement; and (2) select the best segment of spectrum to utilize. The first task is called \emph{spectrum share allocation}, and the second task is called \emph{channel selection}.

\subsubsection{Ecology-inspired spectrum share allocation}
Suppose a TVWS channel is the minimum unit amount of spectrum allocation. Let $A_i$ denote the \emph{amount of spectrum} allocated to network~$i$. Since $N\geq n$, every CR network is assumed to exclusively occupy at least one channel, and thus $A_i \in [1, N-n+1]$ for network~$i\in \mathcal{K}$.

Equivalently, we can rewrite $A_i$ as $A_i = 1 + S_i$, where $S_i \in [0, N-n]$ is the amount of spectrum that is \emph{dynamically} allocated to network~$i$ during the spectrum sharing process. We refer to $S_i$ as the \emph{spectrum share} of network~$i$. Given $n$ competing networks in $\mathcal{K}$,
\begin{itemize}
  \item When $N=n$ (the trivial case), every CR network acquires only one channel, and the dynamically allocated spectrum share is zero.
  \item When $N>n$, every CR network may acquire more than one channel, and the sum of the spectrum share values of all networks is equal to $(N-n)$.
\end{itemize}
Our objective is that the spectrum share allocation process will eventually reach a state of equilibrium, where the spectrum share of each network is proportional to its reported bandwidth requirement.

Spectrum share allocation among the coexisting networks through direct coordination may not be possible (due to a lack of infrastructure), too costly, or may be shunned by the competing network operators because they do not want to provide their network traffic information. Instead of direct coordination, the SHARE framework adopts an indirect coordination mechanism, which is inspired by an interspecific competition model from theoretical ecology.

In ecology, the population dynamics of a species in the interspecific resource competition process can be captured by the L-V competition model. In the context of CR network coexistence, we build a \emph{weighted} competition model to help a CR network to determine the \emph{dynamics of its spectrum share}, given its bandwidth requirement. To complete this task, the mediator exchanges two types of control information with every CR network: (1) network~$i$ reports the current value of $S_i$ to the mediator; and (2) the mediator replies back to network~$i$ with the sanitized data, i.e., sum of spectrum share values of all other coexisting networks, i.e., $\sum_{j\neq i} S_j$.

\subsubsection{Foraging-based channel selection} To maximally fulfill its allocated spectrum share, each CR network is allowed to select up to $\lfloor S_i \rfloor$ channels. Let $\mathcal{C}_i$ denote the set of channels selected by network~$i$, where $ |\mathcal{C}_i| \leq \lfloor S_i \rfloor$. Without explicitly knowing others' selection, it is possible that multiple CR networks select the same channel, thereby causing inter-network interference.

In the optimal foraging theory, an animal is free to move to the patch of resource that has the maximum suitability such that the animal's fitness can be maximized. Similarly, SHARE guides a CR network to always select the channel that has the maximum attractiveness (i.e., the minimum interference).
% As a result, the amount of spectrum for a network can be maximized.

In the channel selection process, two types of control information are exchanged: (1) CR network~$i$ sends $\mathcal{C}_i$ to the mediator; and (2) the mediator calculates the \emph{suitability} of every channel, and sends the values back to network~$i$. Network~$i$ uses this information in its channel selection. The definition of suitability will be given in Section~\ref{sec:foraging}.

\section{An Ecology-inspired Spectrum Share Allocation Algorithm}\label{sec:LV}

\subsection{Weighted-fair Spectrum Share Allocation Problem}
Suppose a set $\mathcal{K}$ of $n$ co-located CR networks have individual bandwidth requirements $R_1,R_2,...,R_n$, and operate over the same TVWS. The first objective for coexisting CR networks is to split the TVWS into $n$ pieces of spectrum shares that are proportional to their individual bandwidth requirements, without sharing individual's bandwidth requirement with each other.

If network~$j$ achieves a spectrum share of $(N-n) R_j/\sum_i R_i$, we say that this spectrum share allocation process is \emph{weighted-fair}.

Let $\mathbf{S}(\mathcal{K})=[S_1, S_2, \dots, S_n]$ denote the \emph{spectrum share vector} for $\mathcal{K}$ over the TVWS\footnote{The vector is a row vector or a $1 \times n$ matrix.}. We define the \emph{fairness index}, $F(\mathbf{S}(\mathcal{K}))$, for networks in $\mathcal{K}$ as follows:
\begin{equation}\label{eqt_weighted_fairness}
F(\mathbf{S}(\mathcal{K})) =\frac{ \left( \sum_{i \in \mathcal{K}} S_i  \right)^2 }{ \sum_{i \in \mathcal{K}} R_i \cdot  \sum_{i \in \mathcal{K}} R_i \left(\frac{S_i} {R_i} \right)^2  }.
\end{equation}
The maximum value of $F(\mathbf{S}(\mathcal{K})) $ is one (the best or weighted-fair case), where the allocated spectrum share value of a network is proportional to its bandwidth requirement.

Let $\mathcal{I}_i$ denote the \emph{set of shared control information} known by network~$i$, and it is easy to see that $R_i \in \mathcal{I}_i$. However, $R_j \notin \mathcal{I}_i$ because competing networks~$i$ and $j$ have conflict of interest issues and customer privacy concerns.

Then, we formulate a \emph{weighted-fair spectrum sharing allocation} problem where heterogeneous CR networks dynamically determine their spectrum share values.
\newtheorem{problem_fair}[problem]{\textbf{Problem}}
\begin{problem_fair}
\label{problem_fair} Given a set of $n$ co-located CR networks, $\mathcal{K}$, operating over $N$ TVWS channels, one has to solve the following problem to find the spectrum share vector for $\mathcal{K}$:
\begin{align*}
 \text{Maximize}  &~~ F(\mathbf{S}(\mathcal{K}))\\
  \text{subject to} &~~ \frac{S_i}{S_j} = \frac{R_i}{ R_j}, R_j \notin \mathcal{I}_i, \forall i,j \in \mathcal{K}.
\end{align*}
\end{problem_fair}
The first constraint $\frac{S_i}{S_j} = \frac{R_i}{R_j}$ guarantees the weighted fairness, and the second constraint implies that a network~$i$ has no idea about any other network~$j$'s bandwidth requirement.

\subsection{A Weighted-fair Spectrum Competition Model}

\subsubsection{The stable equilibrium of the L-V competition model} The L-V competition model provides a method for defining a state of ``stable equilibrium'' and finding the sufficient conditions for achieving it. Consider the interspecific competition process described by equation~(\ref{eqt_LV_Ni}), when $K_i=K_j$ and $\alpha_{ij}=\alpha_{ji}$ for any two species $i$ and $j$, the sufficient condition for stable equilibrium is $\alpha_{ij} <1$.

\subsubsection{The basic spectrum competition model}
In Table~\ref{tab_mapping}, we identify a number of analogies between a biological ecosystem and a CR network system. Based on equation~(\ref{eqt_LV_Ni}) and the analogies, we can easily obtain a \emph{basic} spectrum competition model as follows.
\begin{equation}\label{eqt_LV_Ni_basic}
\frac{d S_i}{d t} = r S_i \left( 1- \frac{S_i + \alpha \sum_{j\neq i} S_j } {N-n} \right),
\end{equation}
where $S_i$ is the spectrum share for network~$i$, and $r$ is an intrinsic rate of increase. In equation~(\ref{eqt_LV_Ni_basic}), the carrying capacity is equal to the sum of spectrum share values of all CR networks, i.e., $N-n$. A competition coefficient $\alpha<1$ will guarantee the stable equilibrium---i.e., all the competing networks will have the same spectrum share value.

Next, we will show how to extend the basic competition model to be a weighted-fair spectrum competition model that complies with the weighted-fairness requirement (i.e., $\frac{S_i}{S_j} = \frac{R_i}{R_j}$ for any two networks~$i$ and $j$) at the stable equilibrium of CR network coexistence.

\subsubsection{The weighted-fair spectrum competition model}
The basic spectrum competition model guarantees a stable equilibrium where all the competing networks have the same spectrum share value. However, solutions to Problem~\ref{problem_fair} must satisfy the requirement of weighted fairness, which implies that the competing networks' spectrum share values are proportional to their bandwidth requirements. For example, if network $i$ has a bandwidth requirement that is twice of that of network $j$, then network $i$'s allocated spectrum share should be twice of the allocated spectrum share of network $j$.

To support the weighted-fairness in spectrum share allocation, we construct a weighted-fair spectrum competition model by introducing the concept of ``sub-species''. We model a CR network as a number of sub-species, and the network with a higher bandwidth requirement would have a greater number of sub-species than a network with a lower bandwidth requirement.

We use the bandwidth requirement $R_i$ as the number of sub-species of network~$i$. Let $S_{i,k}$ denote the spectrum share allocated to the sub-species $k$ of network~$i$, where $k\in [1, R_i]$. In the \emph{weighted} competition model, every sub-species~$k$ of network~$i$ calculates the change in its spectrum share according to the following equation.
\begin{align}\label{eqt_LV_subspecies}
\delta_{i,k} &= \frac{dS_{i,k}}{dt} \nonumber \\
             &=  r S_{i,k} \left( 1- \frac{S_{i,k} + \alpha \sum_{\kappa \neq k} S_{i,\kappa} + \alpha \sum_{j\neq i} S_j  } {N-n} \right).  \nonumber\\
\end{align}
Then, network~$i$ obtains its spectrum share value by combining the spectrum share values of all its sub-species, i.e., $S_i = \sum_{k} S_{i,k}$.

In SHARE, every network~$i$ periodically sends its spectrum share value $S_i$ to the mediator, and then the mediator sends back the sanitized data $\beta_i = \sum_{j\neq i} S_j$ to network~$i$. The spectrum share allocation process terminates when $\delta_{i,k} = 0$ for all $i$ and $k$. Note that the sanitized data $\beta_i$ is useful to address the conflict of interests and privacy issues between competing networks. That is, even though $\beta_i$ is known to network~$i$, it is unable to figure out any other network $j$'s bandwidth requirement, i.e., $R_j \notin \mathcal{I}_i, \forall j\neq i$. The use of sanitized data coincides with the second constraint of Problem~\ref{problem_fair}.

The pseudo-code in~Algorithm~\ref{alg_share} illustrates the spectrum share allocation process, and the detailed steps are described below.
\begin{enumerate}
  \item A CR network~$i$ (which is viewed as a species) starts its spectrum share allocation process by creating a number of $R_i$ sub-species.
  \item At the beginning of every iteration, every sub-species calculates the change rate of its spectrum share (i.e., $\frac{dS_{i,k}}{dt}$) using the sanitized data $\beta_i$ obtained from the mediator.
  \item If the change rate of spectrum share is positive (or negative), a sub-species increases (or decreases) its spectrum share.
  \item At the end of every iteration, send the new spectrum share value to the mediator, and update the value of $\beta_i$ from it.
  \item Last three steps are repeated until there is no sub-species with a non-zero change rate of spectrum share; that is $\frac{dS_{i,k}}{dt}=0$ for every sub-species $k$ of any network~$i$.
  \item The allocated spectrum share for network~$i$ is $\sum_{k} S_{i,k}$.
\end{enumerate}

%%%%%%%%%%%%%%%%%%%%%%%%%%%%%%%%%%%%%%%%%%%%%
% Algorithm 1: Spectrum Share Competition Algorithm
%%%%%%%%%%%%%%%%%%%%%%%%%%%%%%%%%%%%%%%%%%%%%
\begin{algorithm}
\caption{The Spectrum Share Allocation Algorithm.} \label{alg_share}
\algsetup{indent=2em}

\textbf{Input}: the competition coefficient $\alpha$, capacity $N-n$, intrinsic rate of increase $r$, the sanitized data $\beta_i$.

\textbf{Output}: the spectrum share, $S_i$, for network~$i$.

\begin{algorithmic}[1]

\STATE Network~$i$ generates a number of $R_i$ sub-species.
%\STATE Let $S_{i, k}\leftarrow 0, \forall k\in [1, W_i]$.
\STATE Update the value of $\beta_i$ from the mediator.

\WHILE{$\left( \exists k\in[1, R_i], s.t. ~ \delta_{i,k} \neq 0 \right)$}
    \FOR{$k = 1$ to $ R_i $}
%        \STATE Update the value of $\delta_{i,k}$ by equation~(\ref{eqt_LV_subspecies}).
        \IF{$\delta_k \neq 0$}
            \STATE $S_{i,k} = S_{i,k} +\delta_{i,k}$.
        \ENDIF
    \ENDFOR
    \STATE Send $S_i = \sum_{k} S_{i,k}$ to the mediator, and update the value of $\beta_i$.
\ENDWHILE
\STATE $S_i = \sum_{k} S_{i,k}$.
\end{algorithmic}
\end{algorithm}

\subsection{Weighted-fairness and Stability}
In this section, we show the properties of the equilibrium status achieved by Algorithm~\ref{alg_share}. We first prove that the spectrum share allocation algorithm satisfies the requirement of weighted-fairness defined in Problem~\ref{problem_fair}.
\newtheorem{lemma_fairness}[lemma]{\textbf{Lemma}}
\begin{lemma_fairness}
\label{lemma_fairness} Given $n$ coexisting CR networks in $\mathcal{K}$, when $\alpha <1$, the spectrum share allocation process of Algorithm~\ref{alg_share} is weighted-fair in partitioning the TVWS consisting of $(N-n)$ channels.
\end{lemma_fairness}
\begin{proof}
Suppose network~$i \in \mathcal{K}$ has a number of $R_i$ sub-species. The spectrum share allocation problem is equivalent to a problem where all sub-species compete for the resource using the L-V competition model. Since the sufficient condition for the the equilibrium in the L-V competition model, $\alpha <1$, is satisfied, the algorithm will terminates after a finite number of iterations, and all sub-species obtain the same spectrum share at the equilibrium point~\cite{LV_TCP_symbiosis,LV_TCP_ICDCS08}, which is equal to $\frac{N-n}{\sum_{j \in \mathcal{K}} R_j}$. Hence, network~$i$ with $R_i$ sub-species will obtain a spectrum share $R_i \frac{N-n}{\sum_{j \in \mathcal{K}} R_j}$, and thus $\frac{S_i}{S_{i'}} = \frac{R_i}{ R_{i'}}$, $\forall i,i' \in \mathcal{K}$.
\end{proof}

%%%%%%%%%%%%%%%%%%%% sufficient conditions for stable equillibrium.
% to facilitate the spectrum share allocation process. In this model, the competition coefficient is $\alpha$, and thus we can conclude that the sufficient conditions for the stable equilibrium described by equation~(\ref{eqt_LV_subspecies}) is $\alpha<1$.

% Stable means convergence. definition of stability. theorem 1: stability
Then we prove that the equilibrium point achieved by the weighted-fair competition model is stable.
\newtheorem{theorem_stability}[theorem]{\textbf{Theorem}}
\begin{theorem_stability}
\label{theorem_stability} Let $l = \sum_{i\in \mathcal{K}} R_i$ represent the total number of sub-species in the system. The differential equations~(\ref{eqt_LV_subspecies}) describe an $l$-dimensional system where the equilibrium when $S_i = R_i \frac{N-n}{l}$ is stable.
\end{theorem_stability}
\begin{proof}
Suppose CR networks in $\mathcal{K}$ generate a total number of $l$ sub-species. For the sake of simplicity, we assign every sub-species an index from $\{1,...,l\}$. Let $\mathbf{S}^*=[s_1^*,...,s_l^*]$ be the spectrum share vector at the equilibrium point for all sub-species in the system, where $s_i^*$ is the allocated spectrum share of sub-species~$i$ at the equilibrium point. By Lemma~\ref{lemma_fairness}, we have
$s_i^* = \frac{N-n}{l}$, where $i\in [1, l]$, and equation~(\ref{eqt_LV_subspecies}) is equivalent to
\begin{equation}\label{eqt_LV_subspecies_2}
\frac{d s_i^*}{dt} = r s_i^*  \left( 1- \frac{s_i^* + \alpha \sum_{j \neq i, j\in [1, l]} s_j^* }{N-n}  \right) = 0.
\end{equation}
That is, $s_i^* + \alpha \sum_{j \neq i, j\in [1, l]} s_j^* = N-n$.

% http://en.wikipedia.org/wiki/Linearization
We will prove the equilibrium $\mathbf{S}^*$ is stable by linearizing the system equations at this equilibrium point. Let $\mathbf{S}=[s_1,...,s_l]$ be a spectrum share vector for all sub-species at a non-equilibrium point. We denote the differential equation at this point as
\begin{equation}\label{eqt_LV_subspecies_stability}
G_i(\mathbf{S}) = r s_i  \left( 1- \frac{s_i + \alpha \sum_{j \neq i, j \in [1, l]} s_j }{N-n}  \right).
\end{equation}
Let $\Delta s_i= s_i-s_i^*$. By linearizing equation~(\ref{eqt_LV_subspecies_stability}) at the equilibrium point, we obtain
\begin{align}\label{eqt_LV_subspecies_linearization}
G_i(\mathbf{S}) &= G_i({s_1^*,...,s_l^*}) + \sum_{i \in [1, l]} \left( \left. \frac{\partial G_i(\mathbf{S})}{\partial s_i} \right|_{s_1^*,...,s_l^*} \cdot \Delta s_i \right)  \nonumber \\
& = -\left( \frac{r}{l} \right) \Delta s_i - \frac{r\alpha}{l} \sum_{j \neq i, j \in [1, l]} \Delta s_j.
\end{align}
We derive the $l$ by $l$ Jacobian matrix for the above equation~(\ref{eqt_LV_subspecies_linearization}) as follows
\begin{equation*}
A=\left| \begin{array}{ccccc}
 - \frac{r}{l}       & - \frac{r\alpha}{l} & - \frac{r\alpha}{l}  &\ldots    &- \frac{r\alpha}{l} \\
 \\
 - \frac{r\alpha}{l} &- \frac{r}{l}       & - \frac{r\alpha}{l}  &\ldots    &- \frac{r\alpha}{l}\\
 \\
 \vdots             & \ddots              &\ddots                 &\ldots         &\vdots \\
 \\
 - \frac{r\alpha}{l} &- \frac{r\alpha}{l} &\ldots                &- \frac{r\alpha}{l} & - \frac{r}{l}\\
 \end{array} \right|,
\end{equation*}
which is a symmetric matrix. This matrix has two eigenvalues $\lambda = -\frac{r}{l}-\frac{(l-1)r\alpha}{l}$ and $\frac{r(\alpha-1)}{l}$. Since $0<\alpha<1$, the two eigenvalues are negative. Based on the stability theory, the system is stable if all eigenvalues are negative. Hence, the differential equations shown by~(\ref{eqt_LV_subspecies}) describe an $l$-dimensional system and the equilibrium $\mathbf{S}^* = \{s_i^* | s_i^* = \frac{N-n}{l}, \forall i\in[1, l]\}$ is stable.
\end{proof}

\textbf{Convergence time}. Next, we analyze the time required for the proposed algorithm to converge to the stable equilibrium.
\newtheorem{theorem_convergence}[theorem]{\textbf{Theorem}}
\begin{theorem_convergence}
	\label{theorem_convergence} Consider $N$ networks that compete for the same spectrum band, then the time-to-convergence to the SHARE's equilibrium is $T_c= O(\ln(C/l))$.
\end{theorem_convergence}
\begin{proof}
	Similar to the proof of Theorem~\ref{theorem_stability}, there are a total number of $l$ sub-species. Let $A=\sum_{j \neq i, j\in [1, l]} s_j = (l-1)s_0$, and equation~(\ref{eqt_LV_subspecies_2}) can be rewritten as
	\begin{align}\label{eqt_LV_subspecies_3}
	\frac{d s_i}{dt} = r s_i  \left( 1- \frac{s_i + \alpha A}{C}  \right) = 0.
	\end{align}
	By integrating (\ref{eqt_LV_subspecies_3}), we can obtain
	\begin{align}\label{eqt_LV_subspecies_t}
	s_i(t) = \frac{s_0 e^{rt\left(1-\frac{\alpha A}{C}\right)}(C- \alpha A )}{s_0(e^{rt\left(1-\frac{\alpha A}{C}\right)}-1)+(C-\alpha A)}.
	\end{align}
	
	To calculate the time-to-convergence, we consider the time which is required to increase the spectrum share for network~$i$ from $s_0$ to $s_i(t) = s^*=C/l$. By solving (\ref{eqt_LV_subspecies_t}), the time $T_c$ becomes:
	\[
	T_c = \frac{C}{r(C-\alpha A)} \ln \left(\frac{s_i(t)(C-\alpha A-s_0)}{s_0(C-\alpha A-s_i(t))} \right) .
	\]
	The time of convergence of SHARE is $O(\ln(C/l))$, and it is exponentially fast.
\end{proof}

%%%%%%%%%%%%%%%%%%%%%%%%%%%%%%%%%%%%%%%%%%%%%%%%%%%%%%%%%%%%%%%%%%%%%%
%\textbf{Discussion.} Convergence time to the stable equilibrium is not a concern, since this scheme is carried out within the 802.19.1 system interfaces, which can be finished very soon within the wired network.
% Let $\mathcal{A}_i$ denote the set of channels available for use by network~$i$.

%%%%%%%%%%%%%%%%%%%%%%%%%%%%%%%%%%%%%%%%%%%%%%%%%%%%%%%%%%%%%%%%%%%%%%
%%%%%%%%%%%%%%%%%%%%%%%%%%%%%%%%%%%%%%%%%%%%%%%%%%%%%%%%%%%%%%%%%%%%%%
%%%%%%%%%%%%%%%%%%%%%%%%%%%%%%%%%%%%%%%%%%%%%%%%%%%%%%%%%%%%%%%%%%%%%%
% after figured out limit, every network wants to fill the limit:
% 1. w/o coordination, two networks may select/operate on the same one, causing interference.
% 2. w/ indirect coordination, networks can be distributed on different channels autonomously
% the selection result is stable; the selection process will converge and terminate.

%%%%%%%%%%%%%%%%%%%%%%%%%%%%%%
%%%%%%%%%%%%%%%%%%%%%%%%%%%%%%
%%%%%%%%%%%%%%%%%%%%%%%%%%%%%%
% no need to distinguish insufficient and sufficient cases.
% in insufficient case, the system satisfaction is low.
% sufficient case is our main focus.

\section{A Foraging-based Channel Selection Algorithm}\label{sec:foraging}

\subsection{The Channel Selection Problem}

Given the allocated spectrum share, network~$i$ is allowed to select up to $M_i= \lfloor S_i \rfloor +1$ channels. We call $M_i$ as the \emph{number of allocated channels} of network~$i$, i.e., $|\mathcal{C}_i| \leq M_i$.
% In the trivial case when $N=n$, $M_i = 1, \forall i$.
Without direct coordination (e.g., sharing of control information such as $\mathcal{C}_i$ and $\mathcal{C}_j$), it is possible that two networks~$i$ and~$j$ select the same channel, and the resulting inter-network interference will degrade the network performance.

To minimize the interference with other networks, every CR network in $\mathcal{K}$ tries to select channels with the highest quality (or the least interference) so as to maximally utilize its allocated spectrum share. This process is similar to the behaviors of animal in the IFD model: an animal selects a patch of resources that has the highest suitability such that its own fitness can be maximized~\cite{book_foraging}, which leads to an evolutionary stable equilibrium. We have created analogies between the foraging behavior and the channel selection process\footnote{Detailed analyses can be found in \cite{dyspan2018}}, as given in Table~\ref{tab_foraging}.

Based on the mapping between channel selection and IFD processes, we consider the $N$ channels as $N$ disjoint patches of resource in an environment that are indexed by $i = 0, . . . , N-1$. Since every network $i$ is allowed to select up to $M_i$ channels, a CR network will create a number of $M_i$ \emph{network agents} to complete the channel selection task.

\begin{table}
% \hspace{0.033\textwidth}
  \centering
   \caption{\textnormal{A mapping between the animal's foraging behavior and the CR network's channel selection process.}}
  \label{tab_foraging}
%\small
\begin{tabular}{|c||c|}
\hline
\bfseries      \textbf{Foraging}   &  \textbf{Channel selection}             \\
\hline\hline
A patch of resource            & A TVWS channel                   \\
\hline
An animal                      &         A network agent          \\
\hline
Suitability of a patch         &   Selectivity of a channel                \\
\hline
\end{tabular}
%\normalsize
\end{table}

Let $y_i$ denote the amount of agents that selects channel~$i$. The total population of agents is
$ P = \sum_{i\in [0, N-1]} y_i = \sum_{i\in \mathcal{K}} M_i$. These agents are indexed by the mediator as $0, 1,..., P-1$.

% suitability --> selectivity
Similar to the definition of a patch's suitability (equation~(\ref{eqt_suitability})), we define the \emph{selectivity} of a channel $h$ (i.e., channel $h$'s quality) as
\begin{equation}\label{eqt_sel}
e_h = \frac{1}{y_h},
\end{equation}
where $y_h$ is the number of agents that has selected channel $h$.
%%%%%%%%%%%%%%%%%%%%%%%%%%%%%%%%%%%%%%%%%%%%%%%%%%%%%%%%%%
% animal fitness --> agent fitness
We equate the \emph{agent fitness} of an agent that selects channel $h$, $f_h$, to the selectivity of channel $h$. That is, $f_h = e_h$.

Then, we define the system fitness as
\[
\Phi = \min \{ f_0, f_1,..., f_{P-1}\}.
\]
The maximum possible value for $\Phi$ is one. When $\Phi=1$, every network agent exclusively occupies one channel, and every channel is selected by at most one network agent. In other words, network~$i$ occupies a number of $M_i$ allocated channels, and its allocated spectrum share is maximally fulfilled. Then, we formulate the channel selection problem as follows.
\newtheorem{problem_selection}[problem]{\textbf{Problem}}
\begin{problem_selection}
\label{problem_selection}
%Suppose $\mathbf{M}(\mathcal{K}) = [M_1, \ldots, M_n]$ is the vector of number of allocated channels
Given a system of $n$ coexisting CR networks, $\mathcal{K}$, one has to solve the following problem to maximize the system fitness.
\begin{align*}
 \text{Maximize}  &~~ \Phi  \\
  \text{subject to} &~~ \mathcal{C}_j \notin \mathcal{I}_i, \forall i,j \in \mathcal{K}.
\end{align*}
\end{problem_selection}
The constraint $\mathcal{C}_j \notin \mathcal{I}_i$ implies that in the channel selection process, there is no sharing of control information, $\mathcal{C}_i$ and $\mathcal{C}_j$, between any two networks~$i$ and $j$.

\subsection{The Channel Selection Strategy and Algorithm}

%%%%%%%%%%%%%%%%%%%%%%%%%%%%%%%%%
%%%%%%%%%%%%%%%%%%%%%%%%%%%%%%%%%
%%%%%%%%%%%%%%%%%%%%%%%%%%%%%%%%%
% mediator communicate w/ one network, and update fitness value for every channel selection decision.
% 1. net sends req to mediator
% 2. reqs are in a queue
% 3. mediator allows one req to complete its decision
% 4. update fitness, and process next req in the queue.

Similar to the animal's foraging behavior in an IFD process, a network agent under SHARE selects a channel that has the highest selectivity value, and the system tend to reach an equilibrium point where the minimum agent fitness is maximized.

At the beginning of the channel selection process, every network~$i$ knows the number of its allocated channels $M_i$, and the set of its occupied channels $\mathcal{C}_i=\emptyset$. A network starts a channel selection process by sending a request on behalf of one of its agent, and this process terminates until all of its agents have finished the channel selection task. The main procedures of the channel selection process are stated below, and the pseudo-code is given in Algorithm~\ref{alg_select}.
\begin{enumerate}
  \item The mediator processes all received requests sequentially: for a received request from network~$i$, it calculates the channel selectivity values of all channels, and send these values in a response to network~$i$.
  \item The agent of network~$i$ follows a greedy algorithm to select a channel $h$ from $[0, ..., N-1]$, which has the highest selectivity value, i.e.,
      \[
        h=\arg\max_{h\in [0, N-1]} e_h.
      \]
  \item Network~$i$ sends the channel selection decision to the mediator, and channel $h$ is added to $\mathcal{C}_i$
  \item The mediator recalculate the selectivity value of channel $h$ based on the received channel selection decision and equation~(\ref{eqt_sel}), and then continue to process the next request.
\end{enumerate}

%%%%%%%%%%%%%%%%%%%%%%%%%%%%%%%%%%%%%%%%%%%%%
% Algorithm 2: Spectrum selection Algorithm
%%%%%%%%%%%%%%%%%%%%%%%%%%%%%%%%%%%%%%%%%%%%%
\begin{algorithm}
\caption{The Foraging-based Channel Selection Algorithm.} \label{alg_select}
\algsetup{indent=2em}

\textbf{Input}: $M_i$, and $\mathcal{K}$

\textbf{Output}: $\mathcal{C}_i$.

\begin{algorithmic}[1]

\STATE Update $\mathcal{C}_i=\emptyset$.

\WHILE{$|\mathcal{C}_i| < M_i$}
    \STATE Obtain the selectivity values of all channels, $e_h, \forall h\in [0, N-1]$, from the mediator.
    \STATE Select channel $h=\arg\max_{h\in [0, N-1]} e_h$.
    \STATE $\mathcal{C}_i=\mathcal{C}_i \cup \{h\}$.
    \STATE Send the channel selection decision $h$ to the mediator that will recalculate $e_h = \frac{1}{y_h+1}$.
\ENDWHILE
\end{algorithmic}
\end{algorithm}

\subsection{Evolutionary Stable Strategy}
In this section, we use the evolutionary game theory to prove that the above channel selection strategy is indeed an evolutionary stable strategy (ESS). In a game-theoretic perspective, each network agent under SHARE's channel selection strategy is viewed as an individual animal that makes choices among $N$ patches (i.e., $N$ channels) to maximize its fitness according to an IFD process.

Let $P_{\mu}$ denote a population of animals that take strategy $\mu$, and let $f(\mu, P_{\nu})$ be the fitness of an animal that takes strategy $\mu$ in a population of animals that take strategy $\nu$. A strategy $\mu$ is an ESS if both of the following two conditions hold~\cite{book_ESS}.
\begin{enumerate}
    \item For all $\nu \neq \mu$, $f(\nu, P_{\mu})\leq f(\mu, P_{\mu})$.
    \item For all $\nu \neq \mu$, if $f(\nu, P_{\mu}) = f(\mu, P_{\mu})$, then $f(\nu, P_{\omega}) < f(\mu, P_{\omega})$, where $P_{\omega}$ is a population formed from both strategies $\mu$ and $\nu$, and $\omega = q\nu + (1-q)\mu $ for a small $q>0$.
\end{enumerate}

By the following theorem, we establish the relationship between the ESS and the proposed channel selection strategy.
\newtheorem{theorem_ESS}[theorem]{\textbf{Theorem}}
\begin{theorem_ESS}
\label{theorem_ESS} The channel selection strategy of SHARE is an evolutionary stable strategy.
\end{theorem_ESS}
\begin{proof}
First, we consider the channel selection process where network agents make choices among $N$ channels as an IFD process where animals distribute themselves across $N$ patches of resource.

Let $\overline{\mu}$ represent the channel selection strategy of SHARE where an agent always chooses the channel with the highest selectivity to maximize its agent fitness.

Since $\sum_i (S_i+1) = N$, the total population of agents, $P = \sum_i M_i \leq N$. Let $P_{\overline{\mu}}$ represent the population with all agents playing strategy $\overline{\mu}$ such that the IFD is achieved.

Under strategy $\overline{\mu}$, $y_h = 1$ or $0$, and thus the selectivity of a channel $e_h = 1$ or
$\infty$ respectively, $\forall h \in [0, N-1]$. As a result, the strategy is equivalent to a strategy $\overline{\mu}'$ where an agent always chooses a channel $h$ with $y_h=0$. Hence,
\[
f(\overline{\mu}', P_{\overline{\mu}'}) = f_h = 1,
\]
for an agent that selects channel~$h$ in a population of agents using strategy $\overline{\mu}'$.

Suppose that the agent makes a unilateral deviation to strategy $\overline{\nu} \neq \overline{\mu}'$ that corresponds to choosing channel $g \neq h$, where $g\in [0, N-1]$. Since $\overline{\nu} \neq \overline{\mu}'$, $y_g \neq y_h=0$, and then $y_g \geq 1$. Then $f(\overline{\nu}, P_{\overline{\mu}'}) \leq \frac{1}{2} < f(\overline{\mu}', P_{\overline{\mu}'})=f(\overline{\mu}, P_{\overline{\mu}})$.

Since $f(\overline{\nu}, P_{\overline{\mu}}) < f(\overline{\mu}, P_{\overline{\mu}})$, the channel selection strategy of SHARE is an ESS.
\end{proof}
The resulting point of the proposed channel selection strategy $y^* = [y_0^*,..., y_{N-1}^*]$, such that $y_h = 0$ or $1$ for all $h = 0,...,N-1$, is a unique global maximum point that solves Problem~\ref{problem_selection}. This represents that each network agent simultaneously chooses a channel with the highest selectivity (i.e., the least interference) to maximize its own fitness. Any number of simultaneous disturbances from this point will lead to a possible degradation in fitness of agents.

\begin{figure}[t]
\centering
\includegraphics[width=3in]{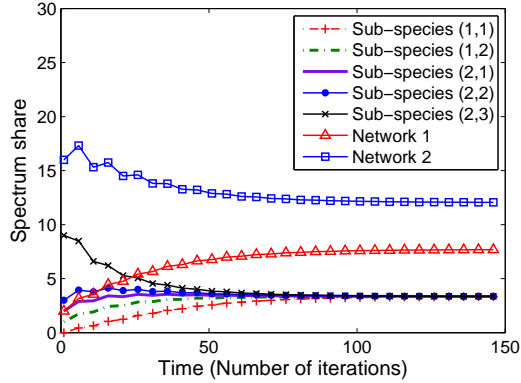}
%\vspace{-0.1in}
\caption{Convergence to the equilibrium.} \label{fig_sim_weighted_LV}
\end{figure}

\begin{figure}[t]
\centering
\includegraphics[width=3in]{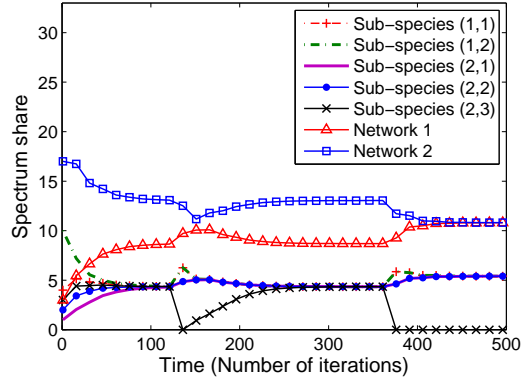}
%\vspace{-0.1in}
\caption{Stability of the equilibrium.} \label{fig_sim_weighted_LV_stable}
\end{figure}

\begin{figure}[t]
\centering
\includegraphics[width=3in]{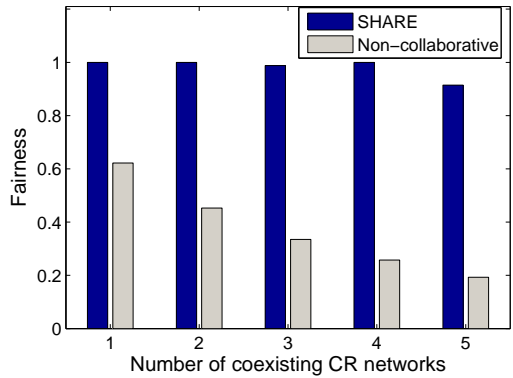}
%\vspace{-0.1in}
\caption{Measured fairness values. } \label{fig_weighted_fairness}
\end{figure}

\section{Performance Evaluation}\label{sec:sim}
In this section, we evaluate the performance of SHARE in two steps. We first investigate the the stability of the equilibrium achieved by the weighted-fair spectrum share allocation scheme. Then, we compare the foraging-based channel selection scheme and the random channel selection strategy in terms of system fitness.

\subsection{The Equilibrium in Spectrum Share Allocation}
In the first set of simulations, we simulate two CR networks under SHARE that coexist over $20$ channels. We fix the bandwidth requirements of two networks as $R_1=2$ and $R_2 =3$, which implies that network\,$1$ has two sub-species and network\,$2$ has three in the spectrum share allocation process using the weighted-fair competition model. In a resource competition process based on the L-V competition model, the chosen parameter values, i.e., the competition coefficient $\alpha<1$ and the intrinsic rate of increase $r<2$~\cite{LV_parameter}, also apply to the proposed model in this paper. The discussions on how to choose appropriate parameters values to achieve the fast convergence to the equilibrium can be found in~\cite{LV_parameter,LV_TCP_symbiosis}, and in this set of simulations we use $\alpha = 0.9$ and $r=1.95$. Next, we show how the coexisting networks under SHARE achieve the equilibrium where the spectrum share of each network is proportional to its bandwidth requirement.

\textbf{Convergence to the equilibrium.} From Figure~\ref{fig_sim_weighted_LV}, we observe the dynamics of the spectrum share value of each network and each sub-species within a network. ``Sub-species $(i,j)$'' in the figure legend represents the sub-species~$j$ within network~$i$. The system converges to the equilibrium state in finite time where all sub-species of every network are allocated the same spectrum share value. The aggregate spectrum share value for all sub-species in a network is proportional to the network's given bandwidth requirement.

\textbf{Stability of the equilibrium.} To test the stability of the equilibrium point, we introduce two types of disturbance in bandwidth requirement by (1) silencing the sub-species\,$(2,3)$ for a short time period, and (2) deleting the sub-species\,$(2,3)$. Figure~\ref{fig_sim_weighted_LV_stable} shows the dynamics of spectrum share values when the disturbance is introduced: even the system is driven away by the change of bandwidth requirement from its current equilibrium status, it quickly converges to a new equilibrium point where the allocated spectrum share values are proportional to the new value of bandwidth requirements.

\textbf{The weighted fairness.} In this simulation, we vary the number of coexisting CR network, and in each simulation run, the bandwidth requirement, $R_i$, for each network~$i$ is randomly chosen from the range $[1, 5]$. Then, we compare SHARE with a ``fair'' allocation scheme that splits the spectrum ``equally'' to $n$ pieces of spectrum share and allocates them to $n$ coexisting networks. Hence, in the fair allocation scheme, all networks get the same spectrum share value regardless of their bandwidth requirements. We measure the weighted fairness values using the fairness index defined in (\ref{eqt_weighted_fairness}). Figure~\ref{fig_weighted_fairness} clearly shows that SHARE allocates spectrum in a weighted-fair manner, and it has an advantage of guaranteeing the high weighted-fairness (close to one).

\subsection{The Channels Selection Strategy}
In this section, we assume the weighted-fair spectrum share allocation scheme, and compare four channel selection strategies: SHARE strategy, ``random'' strategy, and two hybrid strategies. A random strategy prescribes that every network selects a channel $h$ randomly from the set of unoccupied channels ($\{0,1,...,N-1\} \setminus \mathcal{C}$). The first (or second) hybrid strategy is called ``hybrid1'' (or ``hybrid2''), in which only one network (or half of networks) takes the random strategy, while the others follow the SHARE strategy. Besides the system fitness, we use a measure called \emph{collision probability} to define the probability that the collision of channel selection decisions between two networks occurs (i.e., two networks simultaneously select the same channel).

\begin{figure}[t]
\centering
\includegraphics[width=3.3in]{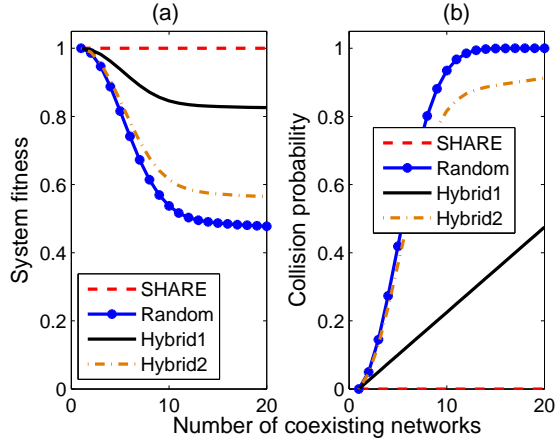}
%\vspace{-0.1in}
\caption{System fitness and collision probability, given various number of coexisting networks. } \label{fig_system_fitness_col}
\end{figure}

\textbf{Number of coexisting networks.} First, we fix the number of allocated channels for every network~$i$ as $M_i=1$, and vary the number of coexisting networks. From simulation results in Figure~\ref{fig_system_fitness_col}, we observe that the system fitness of SHARE is close to one, and other strategies lead to a system fitness much lower than one. Similarly, the SHARE strategy avoids the collision of channel selection decisions of different networks, while other strategies fail to address this problem. For results of either the random or hybrid strategies, we also observe that the system fitness drops (or the collision probability increases) as the number of existing networks increases.

% Five networks
% 20 channels
% Mean # of channels = 4
% variance = 0,1,2,3
% variance up, col down, fitness up.

\textbf{Random bandwidth requirement values.}
We simulate five coexisting networks, and let the bandwidth requirement (BR) for a network~$i$ be a random variable that is uniformly distributed in the range of $[m-\sigma, m+\sigma]$, where $m=4$ and $\sigma=0,1,2,3$. We call $m$ as the mean of BR, and $\sigma$ as the half range of BR's change. In the trivial case when $\sigma=0$, all networks have the same BR value.

\begin{figure}[t]
\centering
\includegraphics[width=3.3in]{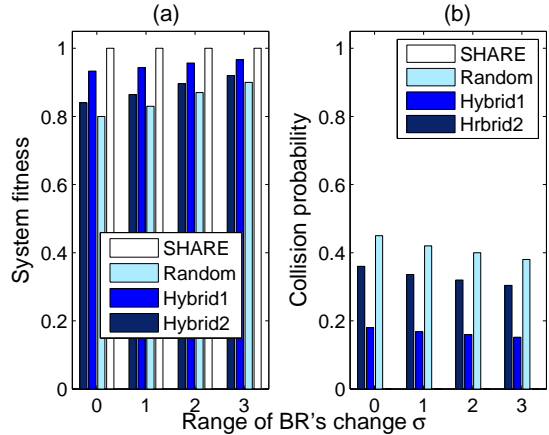}
%\vspace{-0.1in}
\caption{System fitness and collision probability, given random bandwidth requirements. } \label{fig_system_fitness_col_rand_BR}
\end{figure}

According to the weighted-fair spectrum share allocation scheme, the increased value of $\sigma$ will introduce the discrepancy in the number of allocated channels to different coexisting networks. Note that two network agents that belong to the same network will not select the same channel. For a network that takes a random strategy, the increased number of agents will reduce the number of instances of collisions between channel selection decisions. Thus, the increase of $\sigma$ will help lower the collision probability, and thus improving the system fitness for random and hybrid strategies (see results of Figure~\ref{fig_system_fitness_col_rand_BR}).

%\textbf{Number of allocated channels.} When coexisting networks have the same bandwidth requirement, and thus they have the same number of allocated channels, $M_i=\lfloor \frac{N-n}{n} \rfloor+1$ for a network~$i$. We vary the value of $n$.
%\begin{itemize}
%  \item \emph{The case of a small $n$.} When 15 networks coexist, every network~$i$ is allocated two channels, $M_i = 2$. Its available number of channels, $N-I$, is sufficient to satisfy its requirement. In Figure~\ref{fig_satisfaction_n15}, the system fitness under SHARE is close to one given various number of unavailable channels.
%  \item \emph{The case of a large $n$.} In contrast, when three networks coexist and every network~$i$ requires ten channels, $M_i = 10$, its available number of channels, $N-I$, is insufficient to satisfy its requirement. Figure~\ref{fig_satisfaction_n3} shows that the system fitness of both channel selection algorithms drops with the increase of the number of unavailable channels.
%\end{itemize}
%
%

\section{Conclusions}\label{sec:conclude}
Inspired by the symbiotic coexistence in ecology, in this paper we presented a framework called Symbiotic Heterogeneous coexistence ARchitecturE (SHARE), which enables collaborative coexistence among heterogeneous CR networks over TVWS. SHARE enables two heterogeneous CR networks to coexist in TVWS through a mediator-based \emph{indirect} coordination mechanism between them, which avoids the drawbacks of the direct coordination mechanism. Based on the interspecific competition model and the ideal free distribution model in theoretical ecology, we proposed two SHARE algorithms for every coexisting CR network to autonomously complete the two spectrum sharing tasks: (1) dynamically determine its spectrum share that is proportional to its bandwidth requirement, and (2) select channels to improve the system fitness. Analytical and simulation results show that SHARE guarantees a stable equilibrium of coexisting networks in which the weighted-fairness is ensured and the system fitness is maximized.

%\begin{figure}[t]
%\centering
%\includegraphics[width=3in]{satisfaction_rand_workload}
%%\vspace{-0.1in}
%\caption{Random bandwidth requirements given to five coexisting networks.} \label{fig_satisfaction_rand_load}
%\end{figure}

%\theendnotes

%\balancecolumns


\begin{thebibliography}{99}

\small

%\bibitem{LV_sensor_TCP}
%P.~Antoniou and A.~Pitsillides.
%\newblock A Bio-inspired Approach for Streaming Applications in Wireless Sensor Networks based on the Lotka-Volterra Competition Model.
%\newblock {\em Computer Communications}, 33(17):2039--2047, November, 2010.


%
%\bibitem{chan_assign_mobicom1}
%M. Alicherry, R. Bhatia, and L. E. Li.
%\newblock Joint Channel Assignment and Routing for Throughput Optimization in Multi-RadioWireless Mesh Networks.
%\newblock {\em Proc. of ACM Mobicom}, pp. 58--72, September 2005.
%
%
%\bibitem{chan_assign_survey}
%G. Audhya, K. Sinha, S. Ghosh, and B. Sinha.
%\newblock A Survey on the Channel Assignment Problem in Wireless Networks.
%\newblock {\em Wireless Communications and Mobile Computing}, doi: 10.1002/wcm.898, March 2010.


\bibitem{whitefi_sigcomm09}
P.~Bahl, R.~Chandra, T.~Moscibroda, R.~Murty, and M. Welsh.
\newblock White Space Networking with Wi-Fi like Connectivity.
\newblock {\em ACM Sigcomm}, 27--38, Aug. 2009.


\bibitem{tccn2016}
Bhattarai, S., Park, J. M. J., Gao, B., Bian, K., and Lehr, W. 
\newblock An overview of dynamic spectrum sharing: Ongoing initiatives, challenges, and a roadmap for future research. 
\newblock {\em IEEE Transactions on Cognitive Communications and Networking}, 110-128, Jun. 2016.

%
%\bibitem{802191_overview_doc}
%T. Baykas, M. Cummings, H. Kang, M. Kasslin, J. Kwak, Richard Paine, A. Reznik, R. Saeed, and S. J. Shellhammer.
%\newblock Developing a Standard for TV White Space Coexistence: Technical Challenges and Solution Approaches.
%\newblock \url{www.ieee802.org/19/arc/stds-802-19list/docrnXZz7qdyI.doc}.

\bibitem{ukc2006}
K. Bian, J. Park.
\newblock MAC-layer misbehaviors in multi-hop cognitive radio networks.
\newblock {\em US-Korea Conference on Science, Technology, and Entrepreneurship }, pp. 228-248, 2006.


\bibitem{globecom_13}
K. Bian, J. Park, X. Du, X. Li.
\newblock Ecology-Inspired Coexistence of Heterogeneous Wireless Networks.
\newblock {\em IEEE Globecom}, pp. 4921--4926, Nov. 2013.



\bibitem{80222_globe_game}
S. Brahma and M. Chatterjee.
\newblock Mitigating Self-Interference Among IEEE 802.22 Networks: A Game Theoretic Perspective.
\newblock {\em IEEE Globecom}, pp. 1--6, Nov. 2009.

%
%\bibitem{bio_networking}
%F. Dressler and O. B, Akan.
%\newblock Bio-inspired Networking: from Theory to Practice.
%\newblock {\em IEEE Communication Magazine}, 48(11):176--183, November 2010.


\bibitem{Ecma392}
Ecma International.
\newblock ECMA-392: MAC and PHY for Operation in TV White Space.
\newblock 1st Ed., Dec. 2009.

\bibitem{FCC_openTV}
Federal Communications Commission.
\newblock Unlicensed Operation in the TV Broadcast Bands and Additional Spectrum for Unlicensed Devices below 900 MHz in the 3GHz Band.
\newblock ET Docket No. 04-186, May 2004.
%
%\bibitem{LV_econ1}
%R.M. Goodwin.
%\newblock A Growth Cycle.
%\newblock {\em Socialism, Capitalism and Economic Growth, Feinstein}, C.H. Feinstein (ed.), Cambridge University Press, 1967.

\bibitem{IFD}
S. D. Fretwell and H. L. Lucas.
\newblock On territorial behavior and other factors influencing habitat distribution in birds.
\newblock \emph{Acta Biotheor.}, 19(1):16--36, 1970.


\bibitem{80222_odsc}
D. Grandblaise and W. Hu.
\newblock Inter Base Stations Adaptive On Demand Channel Contention for IEEE 802.22 WRAN Self Coexistence
\newblock IEEE docs: IEEE 802.22-07/0024r0, January. 2007.

\bibitem{cellular_Mobicom06}
M. Ghaderi, A. Sridharan, H. Zang, D. Towsley, and R. Cruz.
\newblock TCP-Aware Resource Allocation in CDMA Networks.
\newblock {\em ACM MobiCom}, pp. 215--226, Sept. 2006.

\bibitem{LV_TCP_symbiosis}
G. Hasegawa and M. Murata.
\newblock TCP symbiosis: Congestion control mechanisms of TCP based on Lotka-Volterra competition model.
\newblock {\em ACM Inter-Perf}, Oct. 2006.

%\bibitem{LV_TCP_ICNP07}
%X.M. Huang, C. Lin, F. Ren, G. Yang.
%\newblock Improving the Convergence and Stability of Congestion Control Algorithm.
%\newblock Proc. 15th IEEE Intl. Conf. on Network Protocols.(ICNP 2007). Beijing, China, Oct 2007, pp. 206-215.

\bibitem{LV_TCP_ICDCS08}
X. Huang, F. Ren, G. Yang, and Y. Wu.
\newblock End-to-end Congestion Control for High Speed Networks Based on Population Ecology Models.
\newblock {\em IEEE ICDCS}, pp. 353--360, June 2008.

\bibitem{wifi_zigbee}
J. Huang, G. Xing, G. Zhou, and R. Zhou.
\newblock Beyond Co-existence: Exploiting WiFi White Space for ZigBee Performance Assurance.
\newblock {\em IEEE ICNP}, Oct. 2010.



\bibitem{80211af}
IEEE P802.11 task group af. Wireless LAN in the TV White Space.
\newblock \url{http://www.ieee802.org/11/Reports/tgaf_update.htm/}.


\bibitem{80222}
IEEE 802.22 Working Group.
\newblock \url{http://www.ieee802.org/22/}.

\bibitem{802191}
IEEE 802.19 Task Group 1.
\newblock Wireless Coexistence in the TV White Space.
\newblock \url{http://www.ieee802.org/19/pub/TG1.html}.


%\bibitem{802191_overview_slides}
%IEEE 802.19 Task Group 1.
%\newblock Coexistence in the TV White Space.
%\newblock IEEE doc.: 802.19-10/009r6.
%
%\bibitem{fairness_index}
%R. Jain, D. Chiu, and W. Hawe.
%\newblock A Quantitative Measure Of Fairness And Discrimination For Resource Allocation In Shared Computer Systems.
%\newblock DEC Research Report TR-301, September 1984.
%
%\bibitem{chan_assign_cellular}
%I. Katzela and M. Naghshineh.
%\newblock Channel Assignment Schemes for Cellular Mobile Telecommunication Systems---A Comprehensive
%Survey.
%\newblock {\em IEEE Personal Communications}, 3(3):10--31, June 1996.


\bibitem{8022_TVT_game}
C. Ko and H. Wei.
\newblock Game Theoretical Resource Allocation for Inter-BS Coexistence in IEEE 802.22.
\newblock {\em IEEE Trans. on Vehi. Tech.}, 59(4):1729--1744, May 2010.


\bibitem{LV_model2}
A.J. Lotka.
\newblock Elements of Physical Biology.
\newblock {\em Williams and Wilkins}, 1925.
%
%\bibitem{OFT}
%R. H. MacArthur and E. R. Pianka.
%\newblock On Optimal Use of A Patchy Environment.
%\newblock {\em American Naturalist}, 100:603--609, 1966.

\bibitem{book_ESS}
J. Maynard Smith.
\newblock Evolution and the Theory of Games.
\newblock Cambridge Univ. Press, 1982.

\bibitem{book_foraging}
M. Mingel and C.W. Clark.
\newblock Dynamic Modeling in Behavioral Ecology.
\newblock {\em Princeton Univ. Press}, 1988.


\bibitem{CH_Mobicom06}
A. Mishra, V. Shrivastava, D. Agarwal, S. Banerjee and S. Ganguly.
\newblock Distributed Channel Management in Uncoordinated Wireless Environments.
\newblock {\em ACM MobiCom}, pp. 170--181, Sept. 2006.


\bibitem{LV_parameter}
J.D. Murray.
\newblock {\em Mathematical Biology I: An Introduction}.
\newblock Springer Verlag, 2002.

%
%\bibitem{SenseLess}
%R. Murty, R. Chandra, T. Moscibroda, and P. Bahl.
%\newblock SenseLess: A Database-Driven White Spaces Network.
%\newblock {\em Proc. of IEEE DySpan}, May 2011.
%
%
%\bibitem{LV_econ2}
%G. Nasritdinov, and R.T. Dalimov,
%\newblock Limit Cycle, Trophic Function and the Dynamics of Intersectoral Interaction.
%\newblock {\em Current Research Journal of Economic Theory}, 2(2):32--40, March 2010.
%%
%\bibitem{cellular_IMC10}
%Feng Qian, Zhaoguang Wang, Alex Gerber, Zhuoqing Morley Mao, Subhabrata Sen, and Oliver Spatscheck	 \newblock Characterizing radio resource allocation for 3G networks
%\newblock Proceedin IMC '10 Proceedings of the 10th annual conference on Internet measurement, 2010.

\bibitem{matching_rule}
G. A. Parker and W. J. Sutherland.
\newblock Ideal Free Distribution when Individuals Differ in Competitive Ability: Phenotype-limited Ideal Free Models.
\newblock \emph{Anim. Behav.}, 34(4):1222--1242, Aug. 1986.

\bibitem{cellular_ICNP10}
F. Qian, Z. Wang, A. Gerber, Z.M. Mao, S. Sen, and O. Spatscheck.
\newblock TOP: Tail Optimization Protocol for Cellular Radio Resource Allocation.
\newblock {\em IEEE ICNP}, Oct. 2010.

%\bibitem{IFD}
%N. Quijano and K.M. Passino.
%\newblock The Ideal Free Distribution: Theory and Engineering Application
%\newblock IEEE Transactions on Systems, Man, and Cybernetics, Part B: Cybernetics, 37(1), pp.154--165, Feb. 2007.

%
%\bibitem{chan_assign_NP_hard}
%A. Raniwala and T. Chiueh.
%\newblock Architecture and Algorithms for An IEEE 802.11-based Multi-channel Wireless Mesh Network.
%\newblock {\em Proc. of IEEE INFOCOM}, pp. 2223--2234, March 2005.


\bibitem{80222_globe08}
S. Sengupta, R. Chandramouli, S. Brahma, and M. Chatterjee.
\newblock A Game Theoretic Framework for Distributed Self-Coexistence among IEEE 802.22 Networks.
\newblock {\em IEEE Globecom}, Nov. 2008.

\bibitem{femto_MobiHoc09}
K. Sundaresan and S. Rangarajan.
\newblock Efficient Resource Management in OFDMA Femto Cells.
\newblock {\em ACM MobiHoc}, pp. 33--42, May 2009.

\bibitem{book_coexistence}
M. Tokeshi.
\newblock {\em Species Coexistence: Ecological and Evolutionary Perspectives}.
\newblock Wiley-Blackwell, 1998.
%
%
%\bibitem{chan_assign_mobicom2}
%R. Vedantham, S. Kakumanu, S. Lakshmanan, and R. Sivakumar.
%\newblock Component-based Channel Assignment in Single Radio, Multi-channel Ad Hoc Networks.
%\newblock {\em Proc. of ACM Mobicom}, pp. 378--389, September 2006.



\bibitem{LV_model1}
V. Volterra.
\newblock Variations and Fluctuations of the Number of Individuals in Animal Species Living Together.
\newblock \emph{Animal Ecology}, R.N. Chapman, (ed.), McGraw-Hill, 1931.



\bibitem{femto_Mobicom10}
J. Yun and K. Shin.
\newblock CTRL: A Self-organizing Femtocell Management Architecture for Co-channel Deployment.
\newblock {\em ACM MobiCom}, pp. 61--72, Sept., 2010.

\bibitem{dyspan2018}
Y. Zhang, L. Chen, K. Bian, L. Song and X. Li.
\newblock Enabling Symbiotic Coexistence of Heterogeneous Cognitive Radio Networks.
\newblock {\em IEEE DySPAN}, pp. 1--5, 2018.


\bibitem{zifi}
R. Zhou, Y. Xiong, G. Xing, L. Sun, and J. Ma.
\newblock ZiFi: Wireless LAN Discovery via ZigBee Interference Signatures.
\newblock {\em ACM Mobicom}, pp. 49--60, Sept. 2010.






%\bibitem{chan_assign_WINET}
%S. Ramanathan.
%\newblock A Unified Framework and Algorithm for Channel Assignment in Wireless Networks
%\newblock {\em Wireless Networks}, 5(2):81--94, March 1999.

\end{thebibliography}
\end{document}